\documentclass[reprint,superscriptaddress,nofootinbib,amsmath,amssymb,aps]{revtex4-2}
\usepackage{graphicx}
\usepackage{dcolumn}
\usepackage{bm}
\usepackage{hyperref}
\hypersetup{colorlinks,citecolor=blue,linkcolor=blue,urlcolor=blue,}

\begin{document}
\title{Dynamics of Wave Structures in Multifield Fuzzy Dark Matter Halos}

\author{Yu-Ming Yang}
\email{yangyuming@ihep.ac.cn}
 \affiliation{%
 State Key Laboratory of Particle Astrophysics, Institute of High Energy Physics, Chinese Academy of Sciences, Beijing 100049, China}
\affiliation{
 School of Physical Sciences, University of Chinese Academy of Sciences, Beijing 100049, China 
}%
\author{Xiao-Jun Bi}
\email{bixj@ihep.ac.cn}
\affiliation{%
 State Key Laboratory of Particle Astrophysics, Institute of High Energy Physics, Chinese Academy of Sciences, Beijing 100049, China}
\affiliation{
 School of Physical Sciences, University of Chinese Academy of Sciences, Beijing 100049, China 
}%
\author{Peng-Fei Yin}
\email{yinpf@ihep.ac.cn}
\affiliation{%
 State Key Laboratory of Particle Astrophysics, Institute of High Energy Physics, Chinese Academy of Sciences, Beijing 100049, China}

\begin{abstract}
As a natural extension of the single-field fuzzy dark matter (FDM) model, multifield FDM has attracted increasing attention in recent years. This scenario is motivated both by the axiverse scenario predicted by string theory and by the possibility that multifield FDM may provide a better match to astrophysical observations than its single-field counterpart. In this work, we perform high-resolution numerical simulations to systematically investigate the dynamics of wave structures in multifield FDM halos. In particular, we study the oscillatory and stochastic motions of the central core, the evolution and statistical properties of granules, and the resulting dynamical heating of embedded stellar systems. We find that the frequency spectra of the core density oscillations develop multiple peaks and shift toward higher frequencies relative to the single-field case. The centers of different field components undergo nearly synchronized random walks, while subdominant components exhibit larger random-walk amplitudes. We further show that the suppression of granule density fluctuations with increasing number of fields is largely insensitive to the fractional abundance of each component over a broad parameter range. Moreover, using self-consistent simulations, we find that the dynamical heating induced by granules is progressively suppressed as the number of fields increases. However, once the contribution from the central core is taken into account, this trend would become much less pronounced. 
\end{abstract}

\maketitle
\section{Introduction}
Dark matter constitutes about five times the mass of ordinary baryonic matter in the Universe, yet its particle nature remains completely unknown \cite{1970ApJ...159..379R, Cirelli:2024ssz}. One compelling possibility is that dark matter consists of ultralight bosons with masses around $10^{-22}$ eV, commonly referred to as fuzzy dark matter (FDM) \cite{hu2000fuzzy, peebles2000fluid, hui2017ultralight, hui2021wave, eberhardt2025ultralightfuzzydarkmatter, schive2025fuzzydarkmattersimulations}. Owing to their extremely small masses, these particles possess de Broglie wavelengths on the order of kiloparsecs in galactic environments, giving rise to wave-like behavior on astrophysical scales. As a result, FDM has the potential to alleviate several small-scale challenges faced by the standard cold dark matter paradigm \cite{Bullock:2017xww}.

Most previous studies of FDM have focused on the simplest scenario involving a single ultralight scalar field. From a theoretical perspective, however, a multifield extension is arguably more natural \cite{hui2017ultralight}. A primary motivation for such a scenario comes from string theory, where the compactification of extra dimensions generically predicts a large number of axion-like fields spanning a broad mass spectrum, a framework commonly referred to as the axiverse \cite{Arvanitaki_2010, Marsh_2016}. If multiple axion-like particles have masses in the FDM regime, they could collectively contribute to the dark matter abundance, thereby naturally realizing a multifield FDM scenario. Beyond its strong theoretical motivation, introducing multiple fields also expands the model's parameter space, offering additional flexibility and potentially alleviating some of the stringent constraints imposed on the single-field scenario. These include constraints derived from the Lyman-$\alpha$ forest \cite{Armengaud_2017, Ir_i__2017, Murgia_2018, Rogers_2021}, the (sub)halo mass function \cite{Schutz_2020, Nadler_2021}, and observations of individual dwarf galaxies \cite{Hayashi_2021, Zimmermann:2024xvd, Marsh:2018zyw, Dalal:2022rmp, Teodori:2025rul, May:2025ppj, caputo2026influencetidesselfgravityultralight}.

Over the past few years, several studies have investigated various aspects of multifield FDM \cite{Glennon_2023, T_llez_Tovar_2022, Street_2022, Luu_2020, Pozo_2025, Pozo:2023zmx, Luu_2025, Huang_2023, Eby_2020, Luu_2023, Gosenca_2023, van_Dissel_2024, lehmann2026tunnelingtidalstrippingmultifield}. For example, Ref.~\cite{Luu_2025} performed cosmological simulations and showed that different dark matter halos can host disparate relative abundances of individual fields, thereby naturally leading to a diversity in halo properties. Ref.~\cite{Huang_2023} studied the two-field scenario through cosmological simulations and found that each field can develop its own central core within a given halo \cite{Luu_2023}, with the two cores moving together and remaining approximately concentric. However, when the more massive component contributes only a small fraction of the total dark matter density, its associated core can deviate substantially from the standard soliton profile. Furthermore, Ref.~\cite{Gosenca_2023} demonstrated that increasing the number of fields suppresses granule-induced density fluctuations within halos, suggesting that the resulting dynamical heating of embedded stellar systems may also be reduced. Other works have explored additional aspects of multifield FDM halos, including the core-halo mass relation \cite{van_Dissel_2024} and tidal stripping processes \cite{lehmann2026tunnelingtidalstrippingmultifield}.

In this work, we carry out a comprehensive numerical investigation of wave structures in multifield FDM halos. Our simulations begin with the formation of isolated halos through soliton mergers, which are subsequently used to explore the dynamical evolution of the central cores and the surrounding granules. We show that the frequency spectra of core density oscillations differ markedly from those in the single-field case, displaying multiple characteristic peaks and a systematic shift toward higher frequencies. We confirm the findings of previous studies that the cores associated with different field components undergo highly correlated random walks \cite{Huang_2023}, and further find that less abundant components exhibit larger positional fluctuations. To quantify granule fluctuations, we compute both the power spectrum and the two-point correlation function, confirming the previously reported inverse scaling of granule fluctuation strength with the number of fields \cite{Gosenca_2023} and further finding that this scaling is largely insensitive to the fractional abundance of individual components. Finally, by combining the FDM simulations with $N$-body simulations of stellar systems, we demonstrate that granule-induced dynamical heating becomes progressively weaker as the number of fields increases. However, this dependence becomes much less pronounced once the contribution from the central core is taken into account.

This paper is organized as follows. In Sec.~\ref{Sec_2}, we present the equations of motion governing multifield FDM, and describe the simulation setup and the procedure for constructing isolated multifield FDM halos through soliton mergers. Sections~\ref{Sec_4} and \ref{Sec_5} focus on the dynamical evolution of central cores and the statistical properties of the surrounding granules, respectively. In Sec.~\ref{Sec_6}, we combine the FDM simulations with $N$-body simulations of stellar systems to investigate the resulting dynamical heating. Finally, Sec.~\ref{Sec_7} summarizes our main conclusions.

\section{Halo construction\label{Sec_2}}


Similar to the single-field case, in multifield FDM, the occupation number of each field component within its own de Broglie volume is much greater than unity. Consequently, each component can be described by a classical field \cite{hui2021wave}, $\psi_i\,(i=1,\cdots, N)$, where $N$ is the number of fields. In the non-relativistic limit, each field obeys an independent Schr$\ddot{\text{o}}$dinger equation \cite{Glennon_2023, Gosenca_2023},
\begin{equation}
    i\hbar\frac{\partial\psi_i}{\partial t}=-\frac{\hbar^2}{2m_i}\nabla^2\psi_i+m_iV\psi_i,
    \label{Schrodinger}
\end{equation}
where $m_i$ is the particle mass of the $i$-th field. The different fields are coupled only through the common gravitational potential $V$, which is sourced by the total mass density via the Poisson equation,
\begin{equation}
    \nabla^2V=4\pi G\sum_{i=1}^N\rho_i=4\pi G\sum_{i=1}^N m_i|\psi_i|^2.
\end{equation}
In the framework considered here, interference occurs only within each individual field component, while different components do not interfere with one another.


In this work, we consider both equal-mass and unequal-mass multifield FDM models. For the equal-mass cases, all field components are assigned the same particle mass of $10^{-22}$ eV. We investigate both two-field and four-field configurations. In the two-field setup, we consider several mass fractions, including the single-field limit $(f_1,f_2)=(0,1)$, as well as the multifield cases $(0.1,0.9), (0.3,0.7)$, and $(0.5, 0.5)$, where $f_i$ denotes the mass ratio of the $i$-th component to the total dark matter mass. In the four-field configuration, we assume equal contributions from all components, i.e., $f_1=f_2=f_3=f_4=0.25$. For the unequal-mass models, we restrict our analysis to the two-field scenario, with particle masses $m_1=10^{-22}$ eV and $m_2=3\times 10^{-22}$ eV. As in the equal-mass case, we consider a range of mass fractions: $(f_1,f_2)=(0.1,0.9)$,  $(0.3,0.7)$, $(0.5,0.5)$,  $(0.7,0.3)$, and $(0.9,0.1)$.

\begin{figure}[htbp]
    \centering    
    \includegraphics[width=\linewidth]{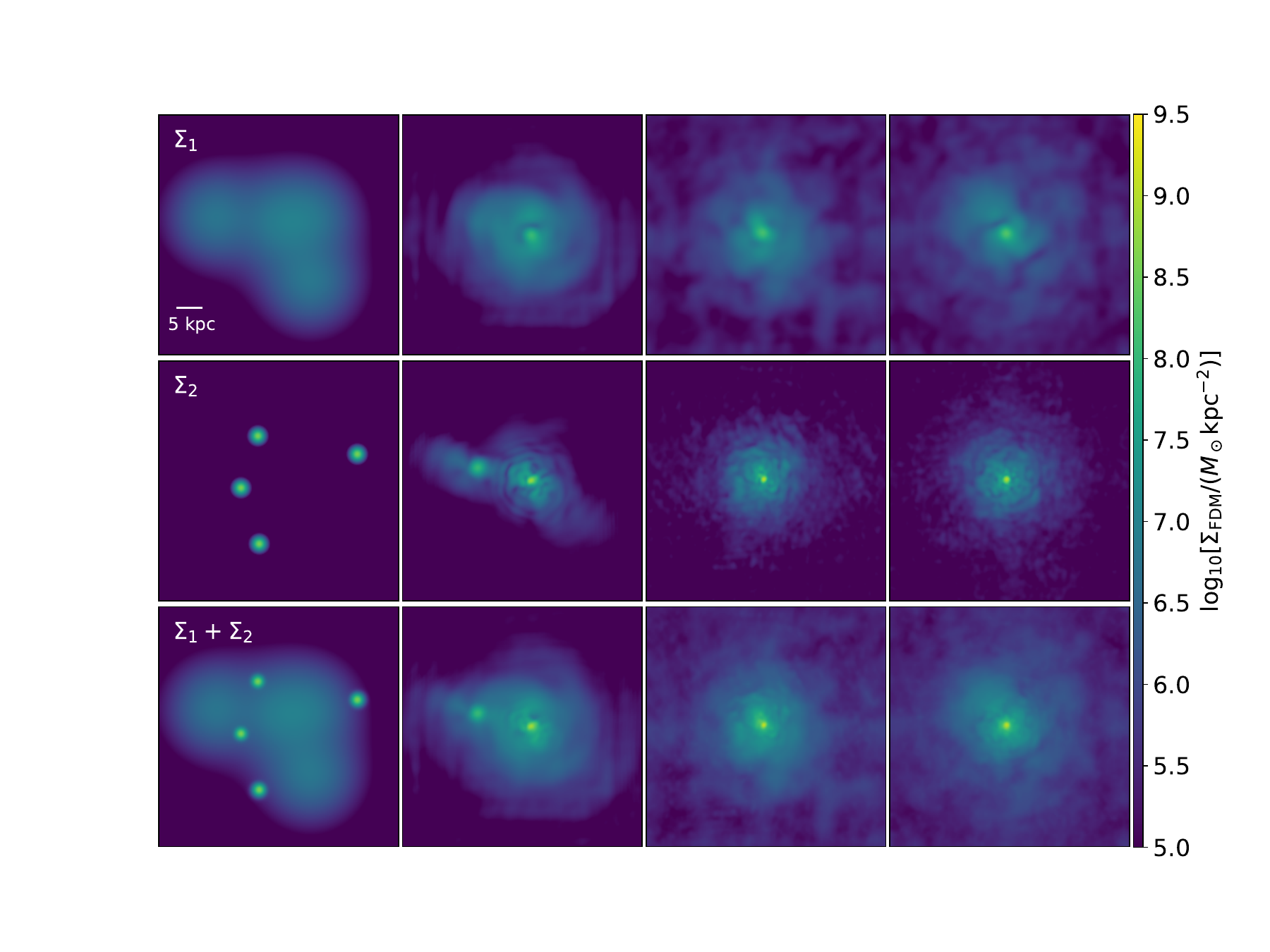}
    \caption{Evolution of the unequal-mass model with $f_1=f_2=0.5$ during the soliton-merger phase, from the initial conditions to 6 Gyr. The top, middle, and bottom rows show the projected density distributions along the $z$-direction of the first field component, the second field component, and the total density, respectively.}
    \label{soliton_merge}
\end{figure} 

We construct isolated, virialized multifield FDM halos through a sequence of soliton mergers. As an illustrative example, consider the unequal-mass model with $f_1=f_2=0.5$. For the first field component, we randomly select four positions within a cubic box of side length 50 kpc. At each position, we place a single-field FDM soliton with the standard soliton density profile \cite{Schive_2014, Schive_2014_1} $\rho_{s}(r)=\rho_c/[1+0.091(r/r_c)^2]^8$, zero initial velocity, and a mass of $M_{s,1}=f_1M_\mathrm{tot}/4$, where $M_\mathrm{tot}$ denotes the total mass of the halo to be constructed. Throughout this work, we adopt a fixed total halo mass of $M_\mathrm{tot}=4\times 10^8\, M_\odot$. The positions of the four solitons are then shifted uniformly such that their center of mass coincides with the center of the simulation box. To minimize the overlap between the initially placed solitons and the box boundaries, we generate multiple random realizations and select one in which all solitons are sufficiently separated from the boundaries. The same procedure is applied to the second field component by placing another four solitons each of mass $M_{s,2}=f_2M_\mathrm{tot}/4$. In this way, the fractions of the two field components are guaranteed to be $f_1$ and $f_2$, respectively. The leftmost panels in the top, middle, and bottom rows of Fig.~\ref{soliton_merge} show the initial projected density distributions along the $z$-direction for the first field component, the second field component, and the total density, respectively. Owing to the larger particle mass of the second field, the corresponding initial solitons are significantly more compact than those of the first field.

Once the initial conditions are constructed, we evolve the system by solving the coupled Schr$\ddot{\text{o}}$dinger equations  Eq.~\ref{Schrodinger}, in which the different field components interact through their common gravitational potential $V$. Under gravitational attraction, the initially separated solitons merge and eventually form a virialized halo. The simulations are performed using the pseudo-spectral method with the \textsc{PyUltraLight} code \cite{Edwards_2018}, which has been extended in this work to support multifield FDM simulations. We adopt a fixed time step of 1 Myr for all simulations. For the equal-mass models, the spatial resolution is set to $256^3$. In contrast, for the unequal-mass models, the resolution is increased to $512^3$ in order to adequately resolve the shorter de Broglie wavelength of the more massive field component. We have verified that the adopted time step and spatial resolution are sufficient to ensure numerical convergence. The pseudo-spectral method naturally implies periodic boundary conditions, which are adopted throughout the simulations.

The temporal evolution of the unequal-mass model with $f_1=f_2=0.5$, from the initial conditions to 6 Gyr, is shown in Fig.~\ref{soliton_merge}. The top, middle, and bottom rows present the projected density distributions along the $z$-direction for the first field component, the second field component, and the total density, respectively. As the initially separated solitons merge under gravity, each field component eventually develops a central core surrounded by granules. Owing to the larger particle mass of the second field, its granules are noticeably smaller than those of the first field, reflecting their shorter de Broglie wavelength. We also find that, when the density fields of the two components are superposed, the fluctuations in the total density are visibly weaker than those in either individual field component. This behavior is consistent with the results of Ref.~\cite{Gosenca_2023}. 

From Fig.~\ref{soliton_merge}, together with a series of additional diagnostics, we find that the halo has reached a sufficiently virialized state after 6 Gyr of evolution. We therefore adopt this snapshot as the initial condition for all subsequent simulations and redefine it as $t=0$. The system is then evolved for a further 10 Gyr, and unless otherwise specified, all analyses presented in the following sections are based on this subsequent evolution.

\section{Core oscillations and random walks\label{Sec_4}}
\begin{figure}[htbp]
    \centering    
    \includegraphics[width=\linewidth]{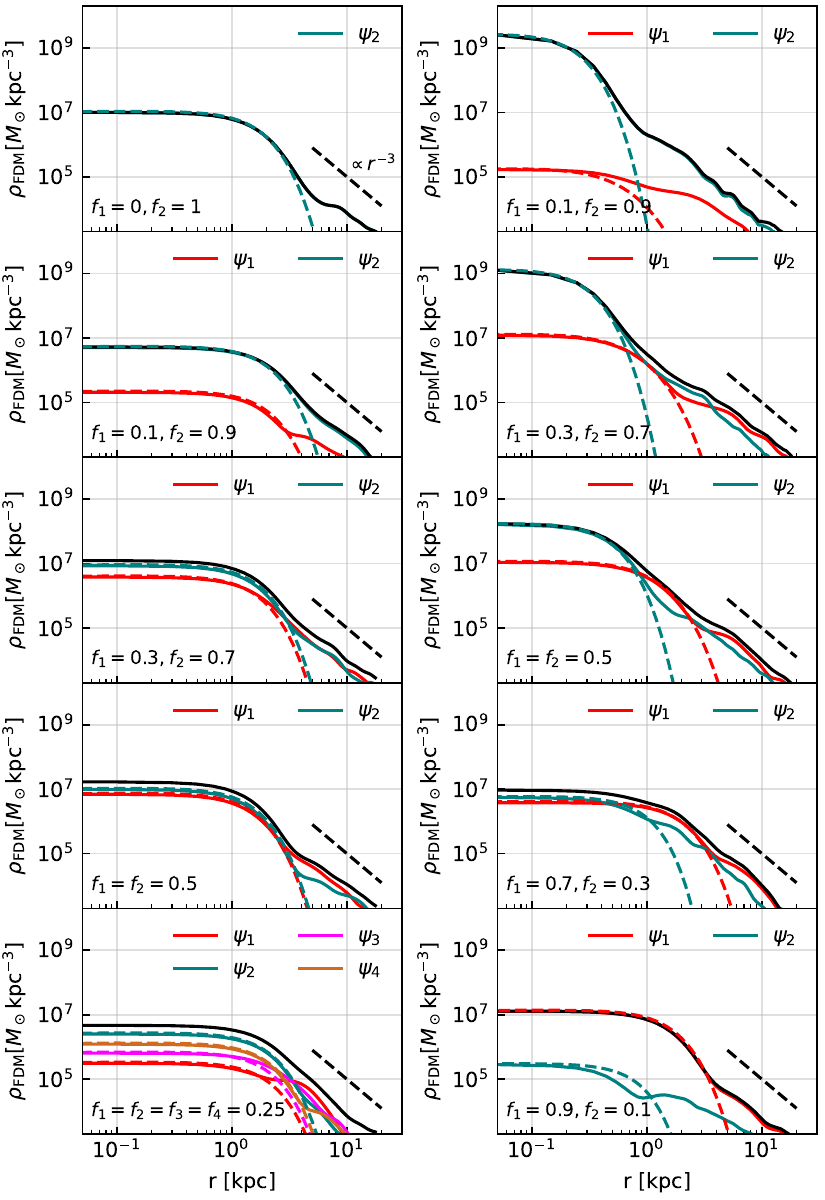}
    \caption{Density profiles at a representative snapshot for the different simulation models. The density profile of each field component is shown as a colored solid line, while the black solid line represents the total density profile. The left and right columns correspond to the equal-mass and unequal-mass models, respectively. The colored dashed lines show the corresponding soliton solutions obtained by numerically solving the stationary Schrödinger equation in the gravitational potential generated by the total density profile.}
    \label{soliton_profile}
\end{figure} 

Refs.~\cite{Huang_2023, Luu_2023} reported that when the more massive field contributes only a sufficiently small fraction of the total dark matter mass, its central core no longer forms a well-defined soliton and instead deviates from the standard soliton profile. In this work, we  examine this conclusion using our simulations. Fig.~\ref{soliton_profile} shows the density profiles at a representative snapshot for the different simulation models. The density profile of each field component is represented by a distinct colored solid line, while the black solid line denotes the total density profile. The left and right columns correspond to the equal-mass and unequal-mass models, respectively. The dashed lines of different colors show the soliton solutions obtained by numerically solving the stationary Schrödinger equation for the ground state in the gravitational potential corresponding to the total density profile (black solid line). For the equal-mass models, we find that the core of each field component is well described by the corresponding soliton solution over the entire range of mass fractions considered. In contrast, for the unequal-mass models, the core of the more massive field begins to show a slight deviation from the soliton solution when its mass fraction decreases to $f_2=0.3$, and the deviation becomes pronounced for $f_2=0.1$. Our results are therefore consistent with the findings of Refs.~\cite{Huang_2023, Luu_2023}.

Fig.~\ref{soliton_profile} also shows that the total density profile in the outer halo is well described by the Navarro-Frenk-White (NFW) profile \cite{Navarro_1996}, exhibiting the characteristic $\rho\propto r^{-3}$ behavior at large radii (black dashed line). Moreover, as the number of field components is increased, the transition from the central soliton-like core to the outer NFW envelope becomes progressively smoother. Such modifications to the inner density profile may leave observable signatures in dwarf galaxy kinematics, such as stellar velocity dispersion profiles. For the equal-mass models, neither the number of fields nor their fractional abundances has a significant impact on the overall density profile. In contrast, the unequal-mass models exhibit substantial diversity in their density profiles as the relative abundances of the two field components are varied \cite{Luu_2025}. Specifically, increasing the mass fraction of the more massive field produces a more compact halo with a denser central region, reflecting the reduced support from quantum pressure associated with the larger particle mass.

\begin{figure}[htbp]
    \centering    
    \includegraphics[width=\linewidth]{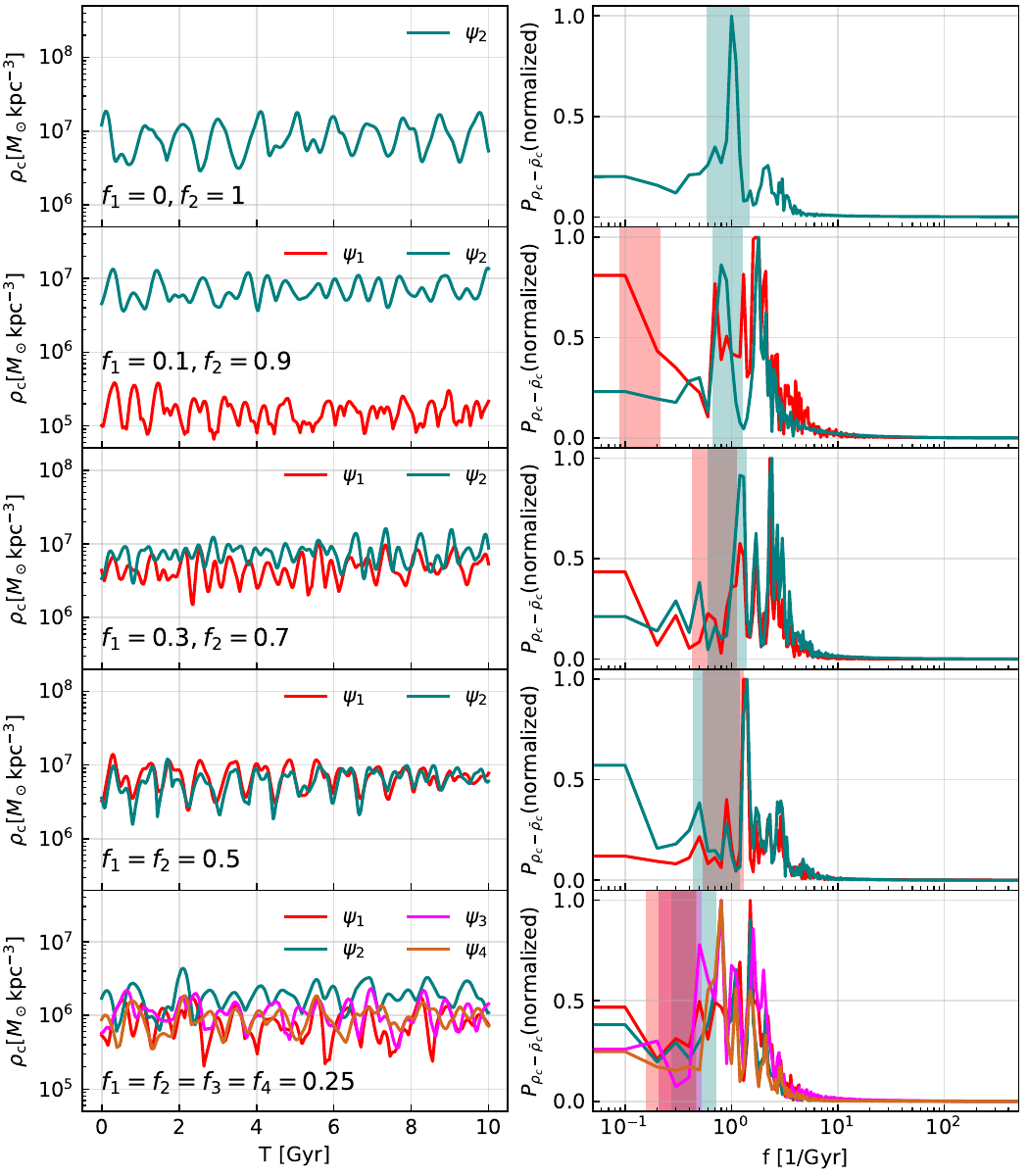}
    \caption{Core oscillations in the equal-mass models. The left column shows the time evolution of the central density of the core for different mass fractions, while the right column presents the corresponding normalized frequency spectra of $\rho_c-\overline{\rho}_c$, where $\overline{\rho}_c$ is the time-averaged central density over the 10 Gyr evolution.}
    \label{core_oscillation}
\end{figure}

\begin{figure}[htbp]
    \centering    
    \includegraphics[width=\linewidth]{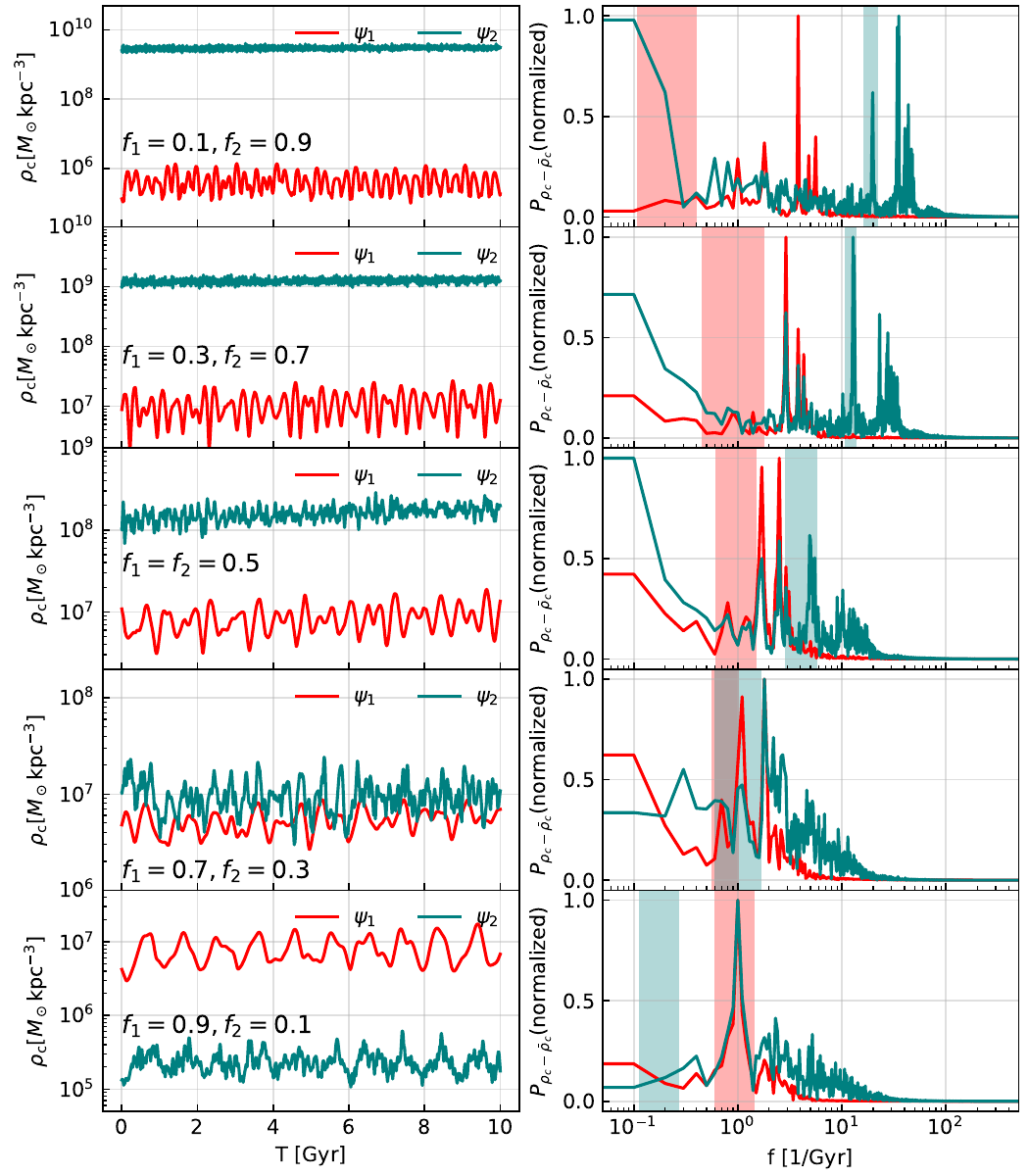}
    \caption{Same as Fig.~\ref{core_oscillation}, but for the unequal-mass models.}
    \label{core_oscillation_1_3}
\end{figure} 

The left column of Fig.~\ref{core_oscillation} shows the time evolution of the central density of the core for the equal-mass models with different mass fractions. The density oscillations of the different field components are found to be largely synchronized, with this synchronization being particularly evident in the case of $f_1=f_2=0.5$. The corresponding normalized frequency spectra of $\rho_c-\overline{\rho}_c$, where $\overline{\rho}_c$ denotes the time-averaged central density over the 10 Gyr evolution, are obtained by performing a Fourier transform, taking the absolute value of the resulting Fourier coefficients, and normalizing the spectra to a maximum value of unity, as shown in the right column. Previous studies have shown that the characteristic oscillation frequency of a single-field FDM core is given by \cite{Veltmaat_2018, Dutta_Chowdhury_2021}
\begin{equation}
    f=10.94\,\mathrm{Gyr}^{-1}\left(\frac{\rho_c}{10^9\,M_\odot\,\mathrm{kpc}^{-3}}\right)^{1/2}.
    \label{frequence}
\end{equation}
Using the minimum and maximum values of $\rho_c$ for each field component over the 10 Gyr evolution, we estimate the corresponding lower and upper bounds on the oscillation frequency predicted by this relation. The resulting frequency ranges are indicated by the shaded regions in the right column of Fig.~\ref{core_oscillation}.

For the single-field case, $(f_1,f_2)=(0,1)$, the frequency spectrum exhibits a single prominent peak, indicating the presence of a well-defined characteristic oscillation frequency, which is in excellent agreement with the frequency range predicted by Eq.~\ref{frequence}. In contrast, for all multifield models except the symmetric case with $f_1=f_2=0.5$, the frequency spectra develop multiple prominent peaks. Moreover, these peaks are systematically shifted toward higher frequencies relative to the range predicted by Eq.~\ref{frequence}, a trend that is also present in the symmetric two-field model. Despite this modification of the spectral structure, the dominant peaks of the individual field components remain approximately aligned in frequency, reflecting the nearly synchronized oscillations of their central cores.

The corresponding results for the unequal-mass models are presented in Fig.~\ref{core_oscillation_1_3}. Similar to the equal-mass case, the gravitational coupling between the two field components gives rise to multiple peaks in the frequency spectra. However, because the second field has a larger particle mass, its intrinsic oscillation frequency is substantially higher than that of the first field. Consequently, when the second field dominates the total mass, the first field is unable to follow its rapid oscillations. This is reflected in the frequency spectra by the fact that the dominant peaks of the second field are shifted toward higher frequencies than those of the first field. The situation changes when the first field dominates the halo. Taking the $(f_1,f_2)=(0.9,0.1)$ case as an example, the gravitational potential is primarily determined by the first field, forcing the second field to oscillate coherently at the same low frequency. At the same time, its intrinsically higher oscillation frequency is not completely erased, but instead manifests itself as a high-frequency modulation superposed on the low-frequency oscillation, as illustrated in the lower-left panel of Fig.~\ref{core_oscillation_1_3}.

\begin{figure}[htbp]
    \centering    
    \includegraphics[width=\linewidth]{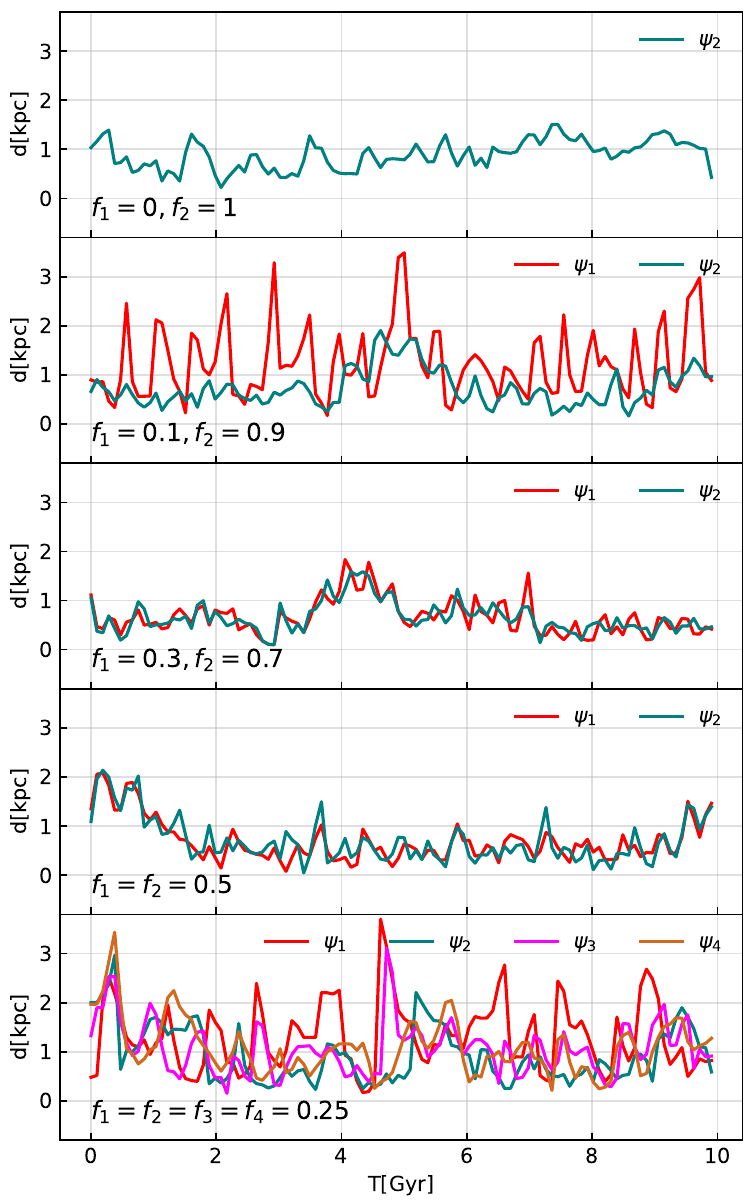}
    \caption{Time evolution of the distance between the density peak of each field component and the center of mass of the entire system for the equal-mass models.}
    \label{core_random_walk}
\end{figure} 
\begin{figure}[htbp]
    \centering    
    \includegraphics[width=\linewidth]{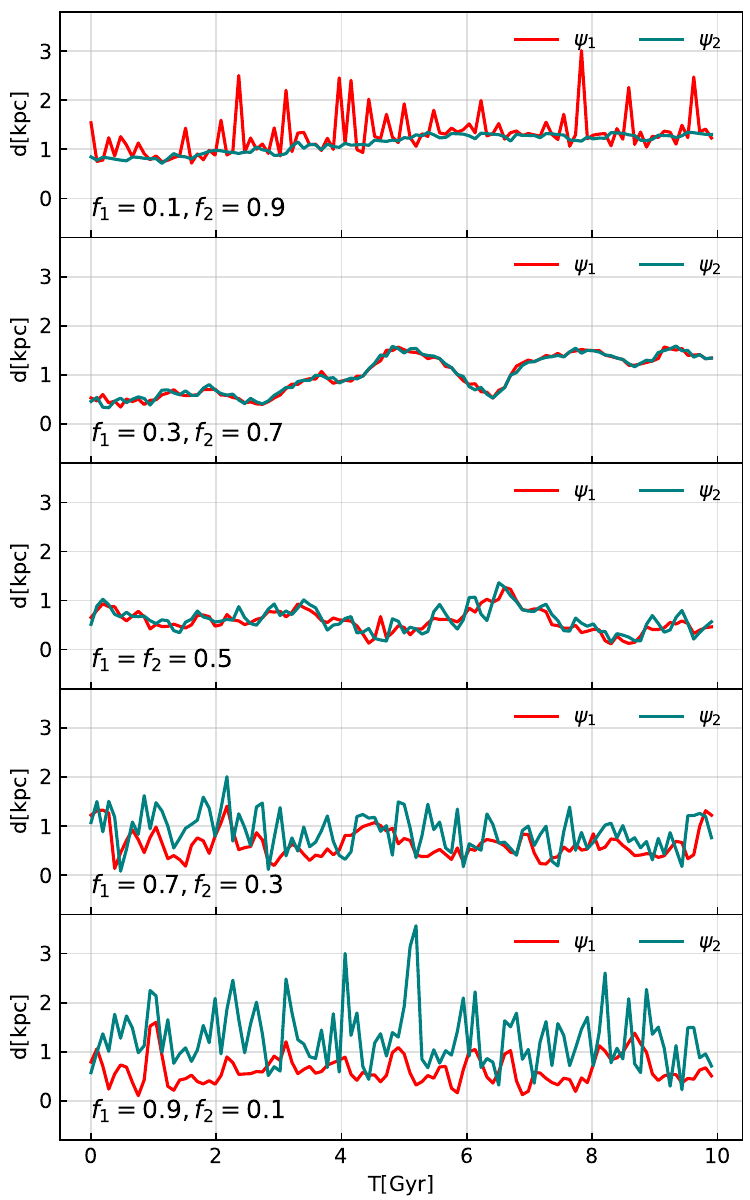}
    \caption{Same as Fig. \ref{core_random_walk}, but for unequal-mass models.}
    \label{core_random_walk_1_3}
\end{figure} 

The time evolution of the distance between the center of each field component (defined as the location of its maximum density) and the center of mass of the entire system is shown in Figs.~\ref{core_random_walk} and \ref{core_random_walk_1_3} for the equal-mass and unequal-mass models, respectively. Owing to their mutual gravitational interaction, the centers of the different field components undergo highly correlated random walks throughout the evolution. Furthermore, when one field contributes only a small fraction of the total halo mass, its core center exhibits substantially larger excursions about the trajectory of the dominant component. This behavior arises because the gravitational potential is primarily determined by the dominant field, causing the subdominant component to behave like a test particle moving in an external gravitational potential well.

\section{Statistical properties of granules\label{Sec_5}}
\subsection{Theoretical description}
In this section, we investigate the statistical properties of granule fluctuations in the NFW region of the halo. For a given density distribution, which may correspond either to an individual field component or to the total density field, we define $\rho(\boldsymbol{x})$ in a coordinate system centered on the density maximum, i.e., the core center. The density fluctuation relative to the spherically averaged density profile $\overline{\rho}(|\boldsymbol{x}|)$ is then defined as
\begin{equation}
    \delta(\boldsymbol{x})\equiv \frac{\rho(\boldsymbol{x})-\overline{\rho}(|\boldsymbol{x}|)}{\overline{\rho}(|\boldsymbol{x}|)},
\end{equation}
following the definition adopted in Refs.~\cite{Gosenca_2023, Amin_2022}. Because the density distribution is smooth within the central core, $\delta(\boldsymbol{x})$ is nearly zero in this region. The quantity $\delta(\boldsymbol{x})$ therefore provides a direct measure of the density fluctuations induced by the granules. One way to quantify these strength of the density fluctuations is through the power spectrum, defined as
\begin{equation}
    P(\boldsymbol{k})\equiv \left|\delta(\boldsymbol{k})\right|^2,
\end{equation}
where 
\begin{equation}
    \delta(\boldsymbol{k})=\frac{1}{\sqrt{V}}\int d^3\boldsymbol{x}\delta(\boldsymbol{x})e^{-i\boldsymbol{k}\cdot\boldsymbol{x}}
\end{equation}
is the Fourier transform of the density fluctuation field $\delta(\boldsymbol{x})$, and $V=(50\,\mathrm{kpc})^3$ is the volume of the simulation box. Another useful statistic is the two-point correlation function, defined as
\begin{equation}
\xi(\boldsymbol{r})\equiv\langle\delta(\boldsymbol{x})\delta(\boldsymbol{x}+\boldsymbol{r})\rangle\equiv\frac{1}{V}\int d^3\boldsymbol{x}\delta(\boldsymbol{x})\delta(\boldsymbol{x}+\boldsymbol{r}).
\end{equation}
The two-point correlation function is related to the power spectrum through a Fourier transform,
\begin{equation}
\xi(\boldsymbol{r})=\int \frac{d^3\boldsymbol{k}}{(2\pi)^3}P(\boldsymbol{k})e^{-i\boldsymbol{k}\cdot\boldsymbol{r}}.
\label{xi_P}
\end{equation}
Rather than the full three-dimensional quantities, the analysis presented below is based on the angular-averaged power spectrum, $P(k)$, and two-point correlation function, $\xi(r)$, defined respectively as
\begin{equation}
    P(k)\equiv\frac{1}{4\pi}\int d\Omega_kP(\boldsymbol{k}),\,\,\,\xi(r)\equiv\frac{1}{4\pi}\int d\Omega_r\xi(\boldsymbol{r}).
\end{equation}

Before presenting the numerical results, it is useful to develop a simple theoretical expectation for the two-field case. Let the total density field be given by $\rho(\boldsymbol{x})=\rho_1(\boldsymbol{x})+\rho_2(\boldsymbol{x})$, where the two components contribute mass fractions $f_1$ and $f_2$, respectively. To leading order, the spherically averaged density profiles of the individual components are expected to satisfy
$\overline{\rho}_1(|\boldsymbol{x}|)/\overline{\rho}(|\boldsymbol{x}|)\simeq f_1$ and $\overline{\rho}_2(|\boldsymbol{x}|)/\overline{\rho}(|\boldsymbol{x}|)\simeq f_2$, where 
$\overline{\rho}(|\boldsymbol{x}|)=\overline{\rho}_1(|\boldsymbol{x}|)+\overline{\rho}_2(|\boldsymbol{x}|)$ denotes the spherically averaged total density profile. The total density fluctuation is therefore given by
\begin{equation}
    \begin{aligned}
        \delta(\boldsymbol{x})&=\frac{\rho(\boldsymbol{x})-\overline{\rho}(|\boldsymbol{x}|)}{\overline{\rho}(|\boldsymbol{x}|)}\\
        &=\frac{\overline{\rho}_1(|\boldsymbol{x}|)}{\overline{\rho}(|\boldsymbol{x}|)}\frac{\rho_1(\boldsymbol{x})-\overline{\rho}_1(|\boldsymbol{x}|)}{\overline{\rho}_1(|\boldsymbol{x}|)}+\frac{\overline{\rho}_2(|\boldsymbol{x}|)}{\overline{\rho}(|\boldsymbol{x}|)}\frac{\rho_2(\boldsymbol{x})-\overline{\rho}_2(|\boldsymbol{x}|)}{\overline{\rho}_2(|\boldsymbol{x}|)}\\
        &\simeq f_1\delta_1(\boldsymbol{x})+f_2\delta_2(\boldsymbol{x}),
    \end{aligned}
\end{equation}
where $\delta_i(x)$ denotes the density fluctuation of the i-th field component relative to its own spherically averaged profile.
We further define the two-point cross-correlation function between the density fluctuations of the two field components as
\begin{equation}
    \xi_{12}(\boldsymbol{r})\equiv\langle\delta_1(\boldsymbol{x})\delta_2(\boldsymbol{x}+\boldsymbol{r})\rangle.
\end{equation}
It follows directly from the definition that $\xi_{21}(\boldsymbol{r})=\xi_{12}(-\boldsymbol{r})$. Consequently, the two-point correlation function of the total density fluctuation can be expressed as
\begin{equation}
    \xi(\boldsymbol{r})\simeq f_1^2\xi_1(\boldsymbol{r})+f_2^2\xi_2(\boldsymbol{r})+f_1f_2\left[\xi_{12}(\boldsymbol{r})+\xi_{12}(-\boldsymbol{r})\right],
\end{equation}
which yields
\begin{equation}
    \xi(r)\simeq f_1^2\xi_1(r)+f_2^2\xi_2(r)+2f_1f_2\xi_{12}(r).
    \label{xi_12}
\end{equation}

In the idealized limit where the two field components have identical particle masses and are statistically uncorrelated, one expects $\xi_1(r)\simeq\xi_2(r)$, $\xi_{12}(r)\simeq 0$, and  $\xi(r) \simeq (f_1^2+f_2^2)\xi_1(r)$. For the special case of equal mass fractions, the amplitude of the density fluctuations is reduced by a factor of two compared to the single-field case $\xi(r) \simeq \xi_1(r)/2$. The above argument can be readily generalized to $N$ statistically independent field components with identical particle masses and equal mass fractions. In this case, the total two-point correlation function becomes 
\begin{equation}
\xi(r)\simeq \frac{\xi_1(r)}{N},
\label{1/N}
\end{equation}i.e., its amplitude is suppressed by a factor of $1/N$ relative to that of a single-field FDM halo. This prediction is consistent with the numerical results reported in Ref.~\cite{Gosenca_2023} and will be further confirmed by our simulations presented below.

Another useful quantity for characterizing the correlation between two field components is the one-point cross-correlation parameter introduced in Ref.~\cite{Gosenca_2023}, which is defined as
\begin{equation}
    \zeta\equiv\frac{\langle\delta_1(\boldsymbol{x})\delta_2(\boldsymbol{x})\rangle}{\sqrt{\langle\delta_1(\boldsymbol{x})^2\rangle}\sqrt{\langle\delta_2(\boldsymbol{x})^2\rangle}}.
\end{equation}
We use this quantity below to quantify the correlation between different field components in our numerical simulations.

\subsection{Simulation results}
\begin{figure}[htbp]
    \centering    
    \includegraphics[width=\linewidth]{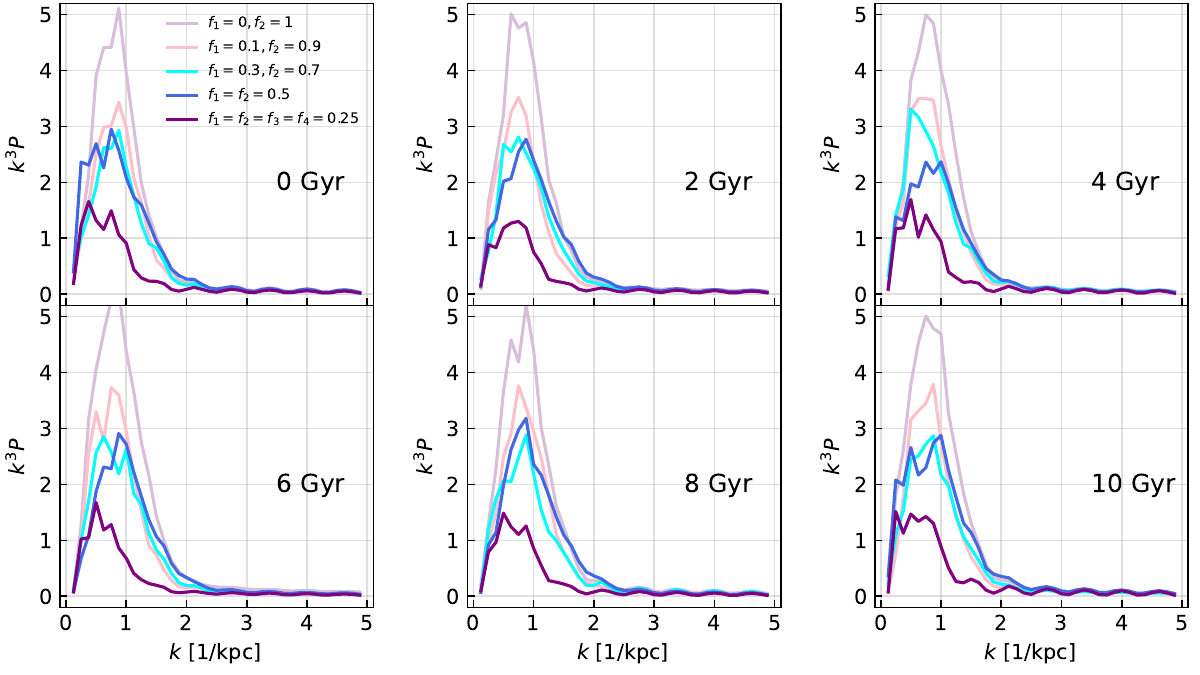}
    \includegraphics[width=\linewidth]{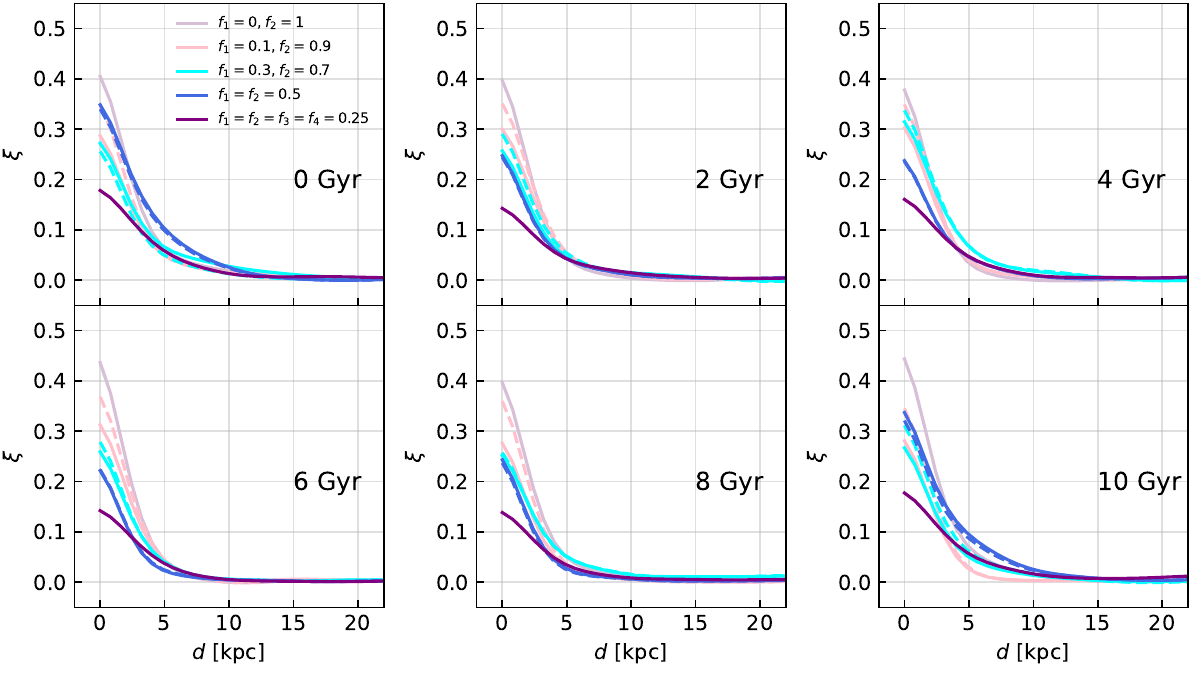}
    \includegraphics[width=\linewidth]{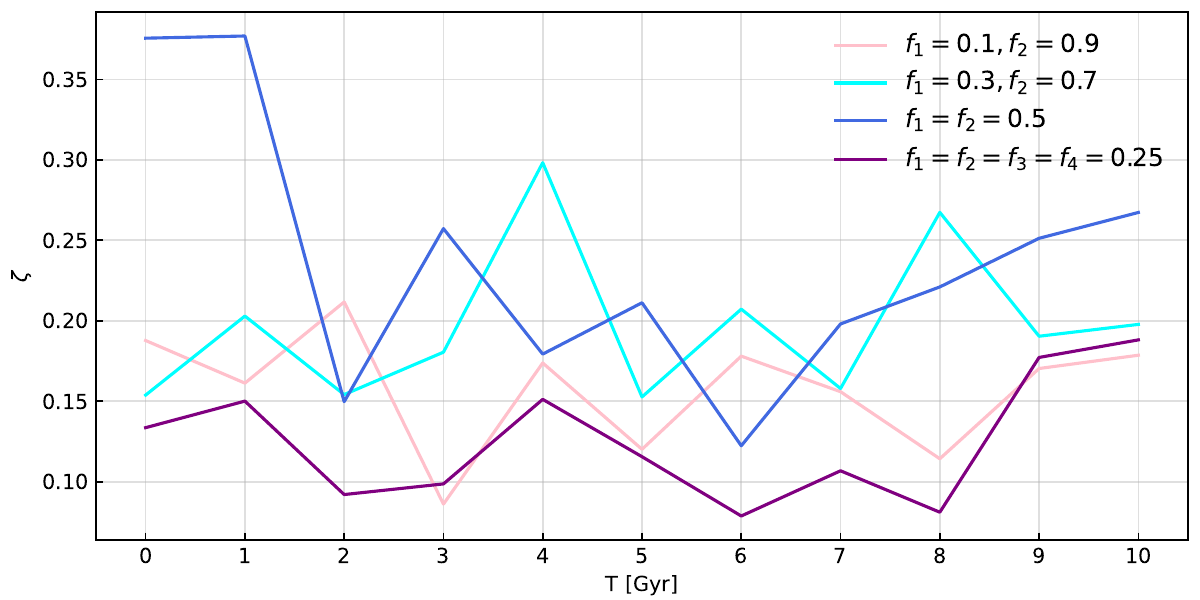}
    \caption{Evolution of the dimensionless power spectrum, $k^3P(k)$ (top six panels), the two-point correlation function, $\xi(r)$ (middle six panels), and the one-point cross-correlation parameter, $\zeta$ (bottom panel), for the equal-mass models. For the four-field model, the one-point cross-correlation parameter is averaged over all pairwise combinations of the four field components.}
    \label{power}
\end{figure} 

Fig.~\ref{power} presents the evolution of the dimensionless power spectrum, $k^3P(k)$, the two-point correlation function, $\xi(r)$, and the one-point cross-correlation parameter, $\zeta$, for the equal-mass models. The top six panels, middle six panels, and bottom panel show these three quantities, respectively. Different colored solid lines correspond to different choices of the mass fractions of the field components. For the four-field model, $\xi(r)$ shown is the average over all pairwise combinations of the four field components.

The peak of the dimensionless power spectrum at $k\sim0.8\,\mathrm{kpc}^{-1}$ reflects the characteristic scale of the density fluctuations, which is set by the de Broglie wavelength of the FDM particles. Since all field components in the equal-mass models have the same particle mass, $m=10^{-22}$ eV, the peak position remains essentially unchanged for different mass fractions.

Focusing on the single-field model and the equal-fraction two- and four-field models, we find that at epochs when the one-point cross-correlation parameter is small, such as $t=2, 4$, and $6$ Gyr, the amplitude of the two-point correlation function decreases systematically with increasing number of fields, approximately following the $1/N$ scaling predicted by Eq.~\ref{1/N}. This behavior indicates that Eq.~\ref{1/N} provides a good description when the correlations between different field components are weak. Since the power spectrum and the two-point correlation function form a Fourier transform pair, as expressed by Eq.~\ref{xi_P}, the peak amplitude of the power spectrum at these epochs also approximately follows the same $1/N$ scaling. However, at epochs when the one-point cross-correlation parameter becomes large, the simple $1/N$ scaling no longer holds. For example, in the equal-fraction two-field model at $t=0$ and $10$ Gyr, the two-point correlation function exhibits a clear deviation from the $1/N$ prediction. In particular, over the separation range of $3-10$ kpc, the two-point correlation function even exceeds that of the single-field model. As we will show below, this behavior arises from the gravitational interaction between the two field components, which induces non-negligible correlations between their density fluctuations.

\begin{figure}[htbp]
    \centering    
    \includegraphics[width=\linewidth]{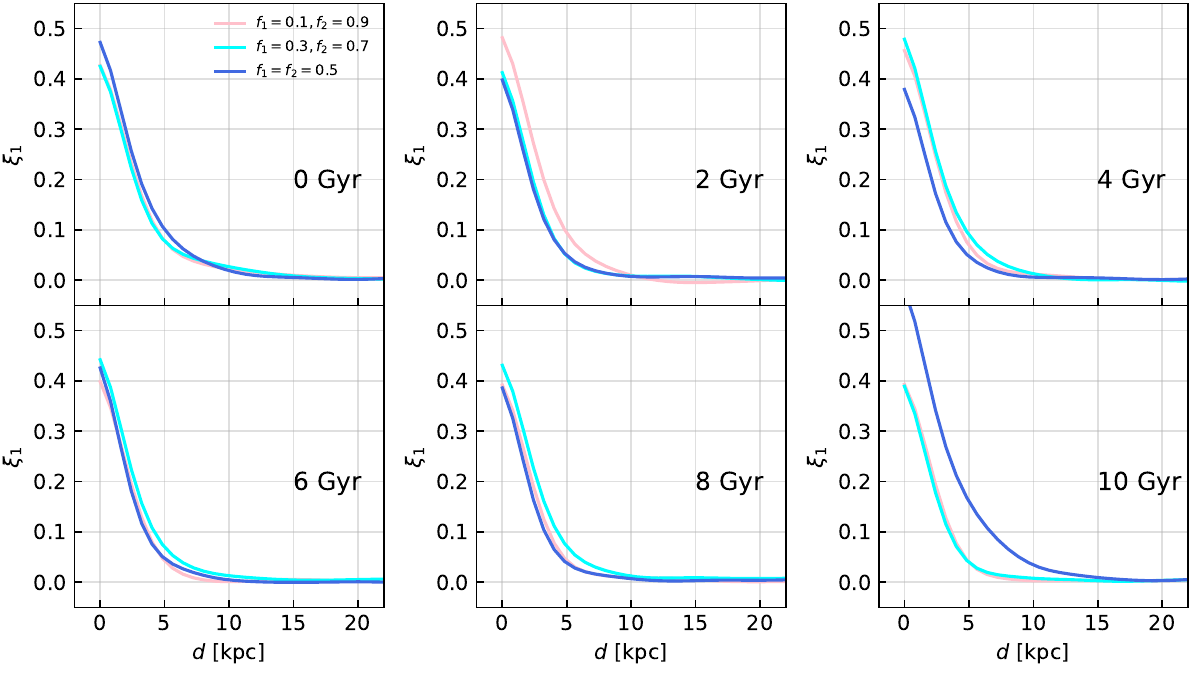}
    \includegraphics[width=\linewidth]{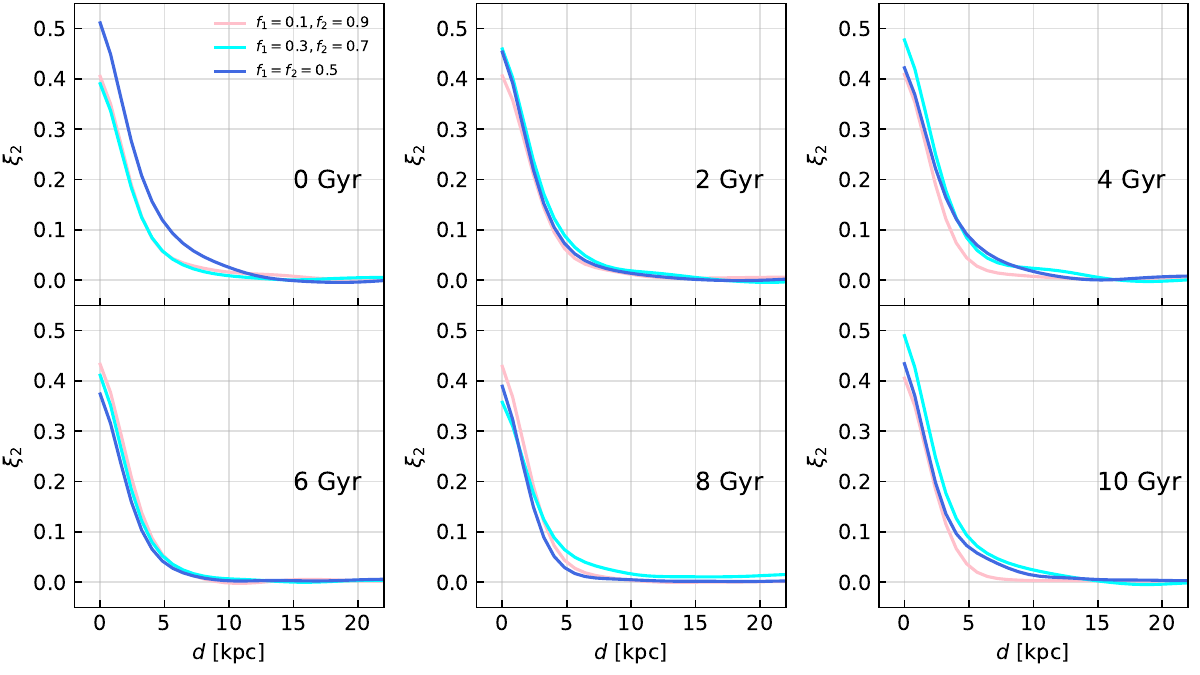}
    \includegraphics[width=\linewidth]{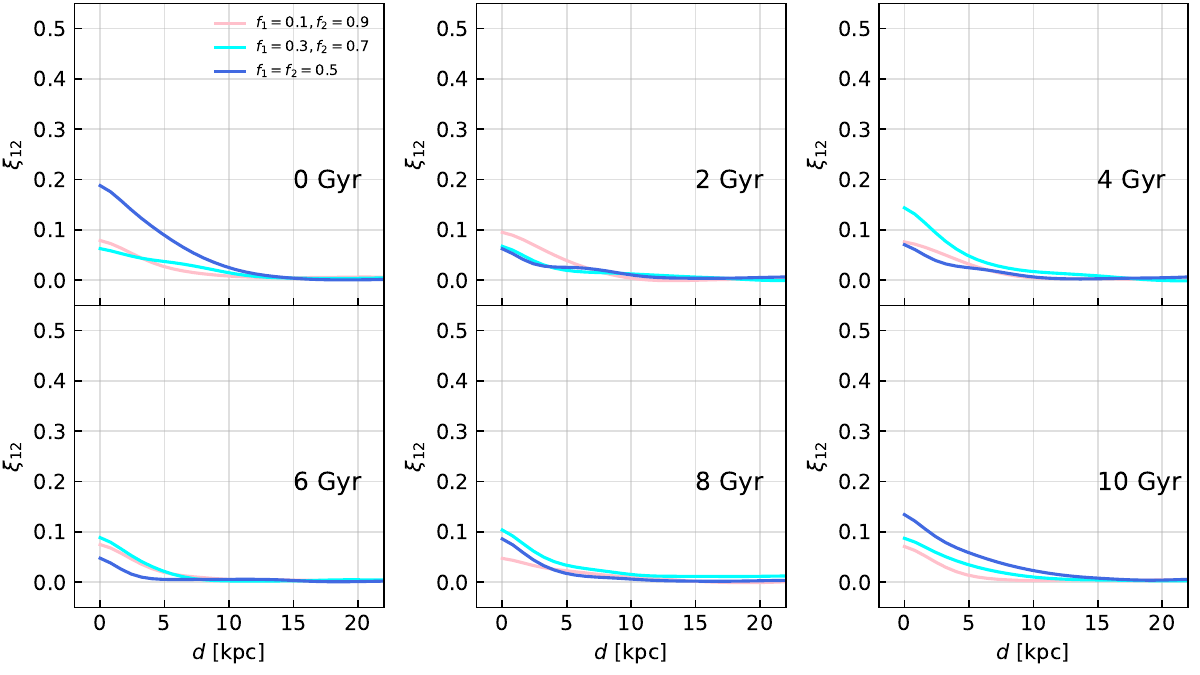}
    \caption{Two-point correlation functions of the individual field components, $\xi_1(r)$ and $\xi_2(r)$, together with the two-point cross-correlation function, $\xi_{12}(r)$, for the equal-mass two-field models at the same epochs as those shown in Fig.~\ref{power}.}
    \label{correlation}
\end{figure} 

To understand the origin of the above behavior, Fig.~\ref{correlation} shows the two-point correlation functions of the individual field components, $\xi_1(r)$ and $\xi_2(r)$, together with the two-point cross-correlation function, $\xi_{12}(r)$, for the two-field models at the same epochs as those shown in Fig.~\ref{power}. We first examine the validity of Eq.~\ref{xi_12}. In the middle six panels of Fig.~\ref{power}, the dashed lines show $\xi (r)$ reconstructed from $\xi_1(r)$, $\xi_2(r)$, and $\xi_{12}(r)$, and the corresponding mass fractions $f_1$ and $f_2$ using Eq.~\ref{xi_12}. The reconstructed lines are in good overall agreement with the direct results, thereby confirming the validity of Eq.~\ref{xi_12}. Discrepancies are present at certain epochs, reflecting the fact that the approximations $\overline{\rho}_1(|\boldsymbol{x}|)/\overline{\rho}(|\boldsymbol{x}|)\simeq f_1$ and $\overline{\rho}_2(|\boldsymbol{x}|)/\overline{\rho}(|\boldsymbol{x}|)\simeq f_2$ are not exact.

Fig.~\ref{correlation} shows that, for different mass fractions in the two-field models, the two-point correlation functions of the individual field components are nearly independent of the mass fractions over most of the evolution and remain close to that of the corresponding single-field model. This indicates that, during most epochs, the mutual gravitational interaction between the two field components is weak and has little effect on the intrinsic granule fluctuations of each field. Noticeable exceptions occur at a few epochs. For example, in the equal-fraction model $f_1=f_2=0.5$, $\xi_2(r)$ at $t=0$ and $\xi_1(r)$ at $t=10$ Gyr deviate significantly from the corresponding results for the other mass fractions, indicating that the gravitational interaction between the two field components becomes much stronger at these epochs. Consistently, both $\xi_{12}(r)$ shown in Fig.~\ref{correlation} and $\zeta$ shown in Fig.~\ref{power} demonstrate that a strong correlation has developed between the two field components. It is precisely this enhanced correlation that causes the total $\xi(r)$ to deviate from the simple $1/N$ scaling.

Fig.~\ref{power} also shows that the correlation between different field components does not increase monotonically with time. Instead, it exhibits a stochastic temporal evolution, becoming strong only during certain epochs. Furthermore, for the two-field models, the mass fraction has only a minor effect on both the power spectrum and the two-point correlation function, at least over the range of mass fractions considered in this work. As a result, the fluctuation amplitudes corresponding to different mass fractions vary irregularly with time, with no systematic ordering among the different models.

\begin{figure}[htbp]
    \centering    
    \includegraphics[width=\linewidth]{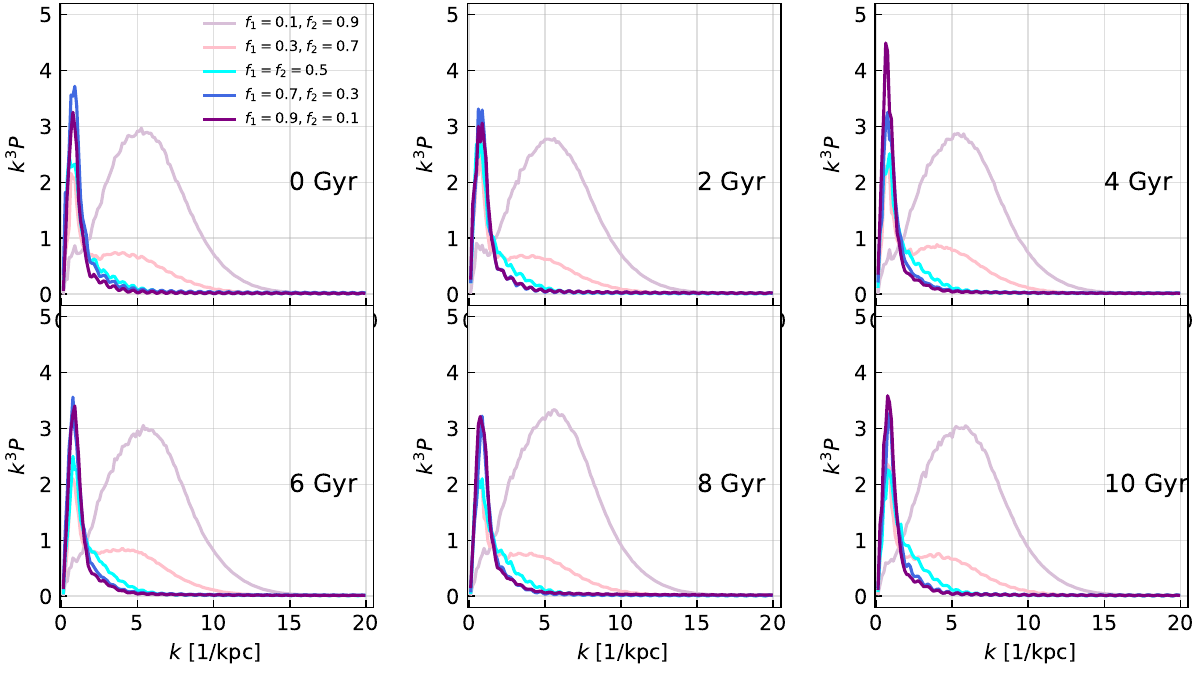}
    \includegraphics[width=\linewidth]{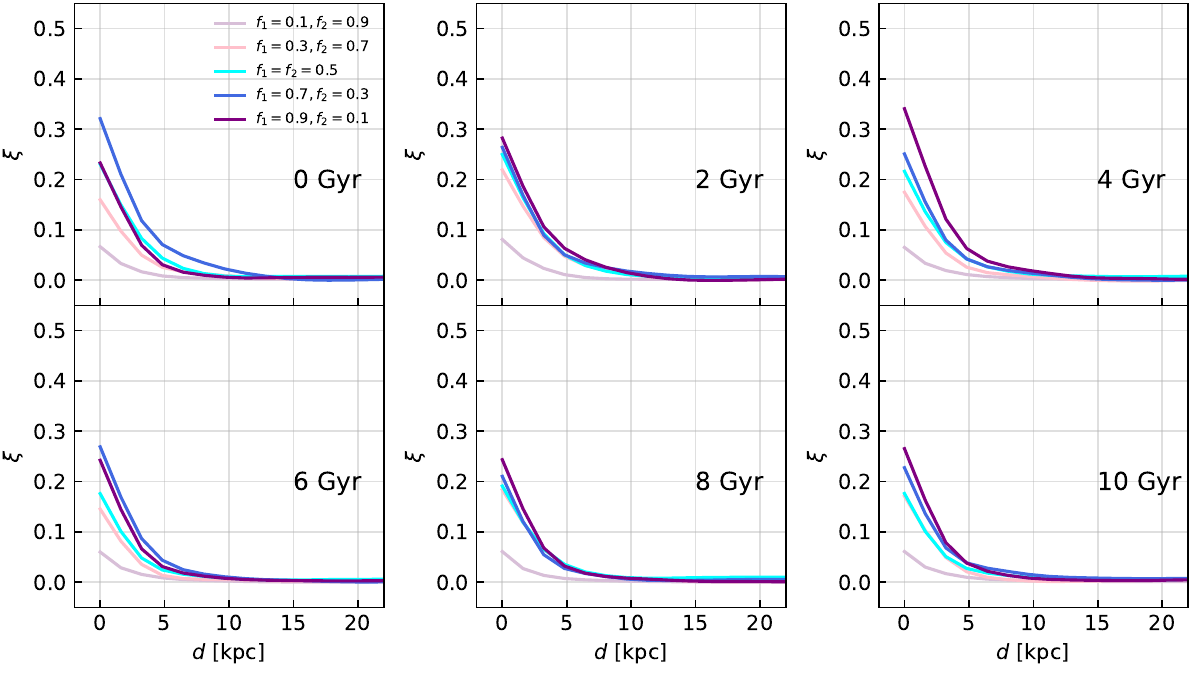}
    \includegraphics[width=\linewidth]{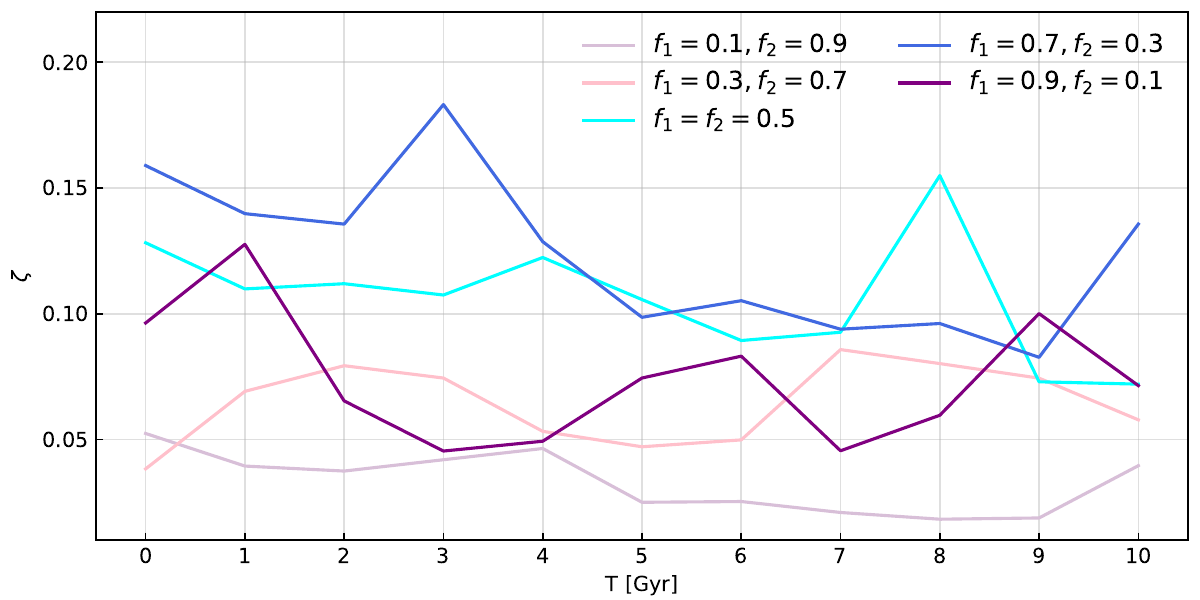}
    \caption{Same as Fig. \ref{power}, but for unequal-mass models.}
    \label{power_1_3}
\end{figure} 

The corresponding results for the unequal-mass models are presented in Fig.~\ref{power_1_3}. When the more massive field component dominates the halo, i.e., for $f_1=0.1$ and $f_2=0.9$, the peak of the power spectrum shifts noticeably toward higher wavenumbers ($k\sim 5\,\mathrm{kpc}^{-1}$), corresponding to smaller spatial scales. This shift reflects the shorter de Broglie wavelength associated with the larger FDM particle mass. Correspondingly, this model also exhibits the smallest amplitude of the two-point correlation function among all the mass fractions considered, indicating the weakest density fluctuations. As the mass fraction of the heavier field decreases to $f_2=0.7$, the main peak of the power spectrum shifts back toward larger spatial scales ($k\sim 1\,\mathrm{kpc}^{-1}$). Nevertheless, $k^3P(k)$ still retains substantial power on small scales ($2-10\,\mathrm{kpc}^{-1}$), indicating that the contribution from the more massive field component to small-scale fluctuations remains significant. As the mass fraction of the heavier field is further reduced to $f_2\leq 0.5$, the small-scale component of the power spectrum almost completely disappears.

\section{Dynamical heating\label{Sec_6}}

In traditional single-field FDM halos, time-dependent density fluctuations arising from the stochastic motion and oscillation of the central soliton, as well as from granule fluctuations, generate gravitational potential fluctuations. These fluctuations can transfer energy to embedded stellar systems over time, causing them to expand and increasing the velocity dispersion of their constituent stars. This process is commonly referred to as dynamical heating \cite{Bar_Or_2019, Dutta_Chowdhury_2023, Yang:2024vgw, Yang_2025}. The analysis presented in the preceding sections shows that increasing the number of field components suppresses the granule-induced density fluctuations, as evidenced by the reduced amplitudes of both the power spectrum and the two-point correlation function. It is therefore reasonable to expect  that the number of field components also affect the efficiency of dynamical heating. In the following, we investigate this effect by combining our multifield FDM simulations with $N$-body simulations of stellar systems. Specifically, we consider the single-field ($f_1=0,f_2=1$), equal-fraction two-field ($f_1=f_2=0.5$), and four-field ($f_1=f_2=f_3=f_4=0.25$) cases within the equal-mass framework.

\begin{figure}[htbp]
    \centering    
    \includegraphics[width=\linewidth]{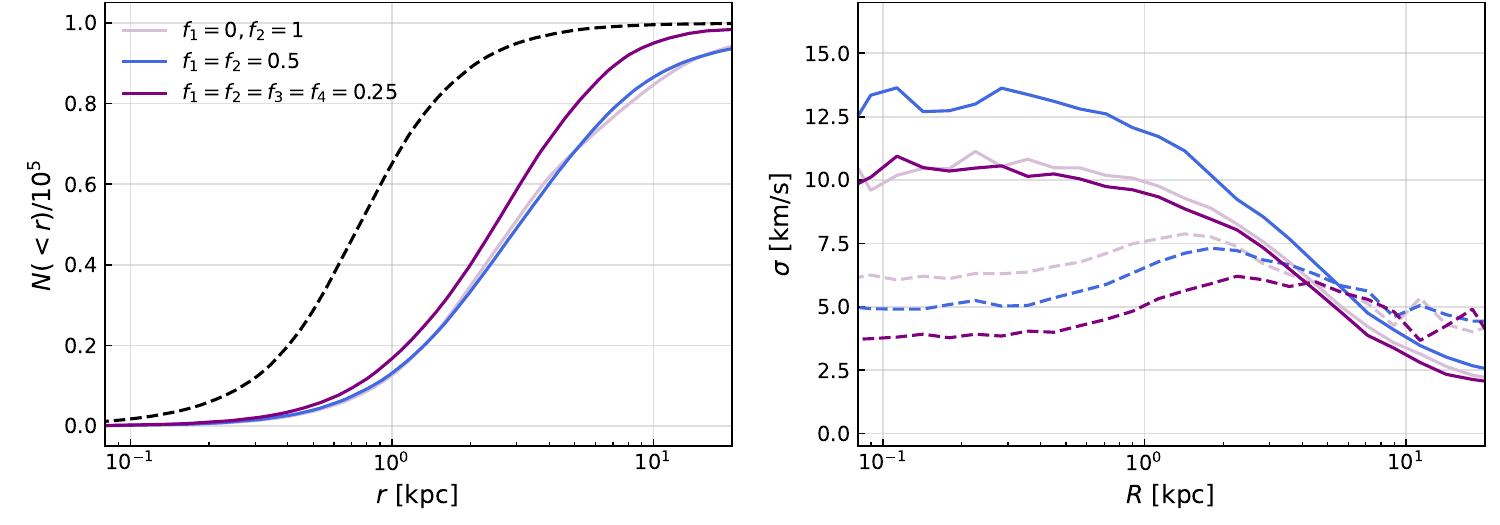}
    \includegraphics[width=\linewidth]{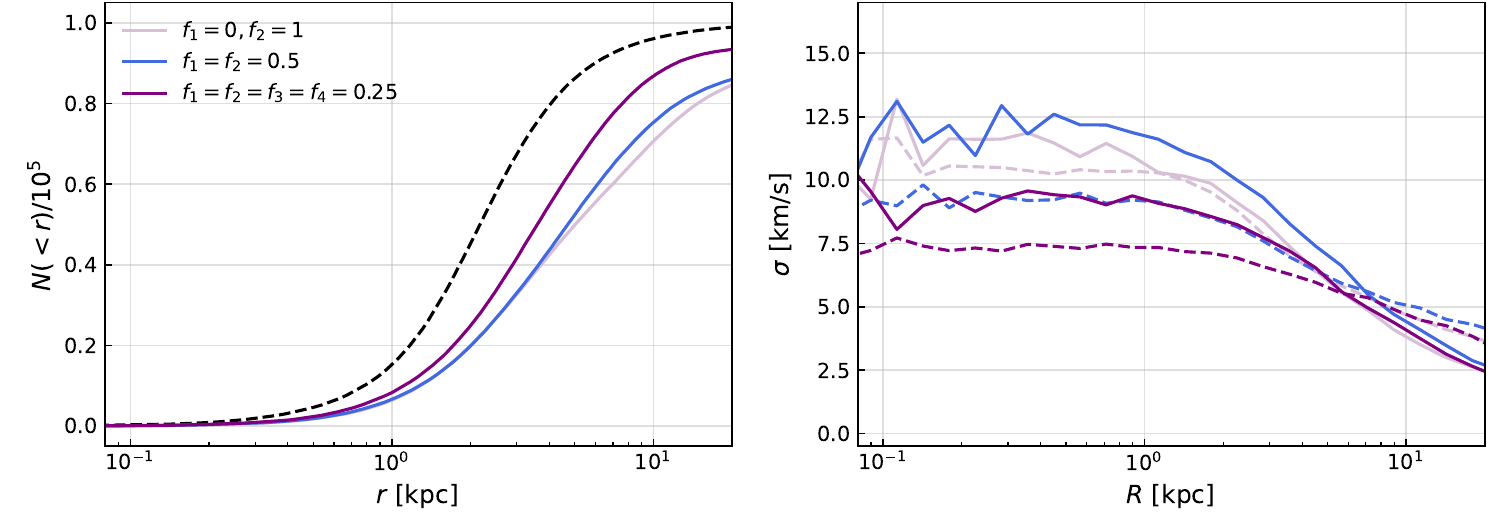}
    \caption{Evolution of the accumulated stellar number distribution (left panels) and the radial profiles of the velocity dispersion along the $z$-direction (right panels). The top and bottom rows correspond to the models with initial stellar scale radii of $0.7$ kpc and $2$ kpc, respectively.}
    \label{heating_soliton}
\end{figure} 

We first consider the case in which the stellar system is initially in virial equilibrium within the gravitational potential of the multifield FDM halo. This requires constructing appropriate initial conditions for the stellar particles. We assume that the initial number density of stars follows a Plummer distribution\cite{Plummer:1911zza},$n(r)\propto (1+r^2/a^2)^{-5/2}$, with two choices of the scale radius, $a=0.7$ kpc and $2$ kpc. The initial positions of the stellar particles are generated by randomly sampling this distribution. We compute the gravitational potential associated with the total density profile of the multifield FDM halo at the beginning of the simulation, i.e., after the $6$ Gyr soliton-merger evolution described in the previous section. The initial velocities of the stellar particles are then sampled from the corresponding distribution function using the Eddington inversion formula\cite{1916MNRAS..76..572E}. This approach ensures the stellar system is in dynamical equilibrium within the halo potential. Each simulation contains $10^5$ stellar particles, which are treated as massless test particles evolving in the time-dependent gravitational potential generated by the multifield FDM. Further details of the set of the initial stellar particles and their subsequent evolution can be found in Ref.~\cite{Yang_2025}.

It should be noted that the initial halo density profiles are not identical for different multifield models. Consequently, the stellar systems constructed through the Eddington inversion naturally possess different initial velocity-dispersion profiles, making a direct comparison between models not entirely straightforward. To minimize this complication, we restrict our analysis to the equal-mass models, for which the halo density profiles are very similar. By contrast, in the unequal-mass models, the halo density profile varies substantially with the mass fractions of the two field components, preventing a meaningful comparison of the dynamical heating efficiency.

The upper-left panel of Fig.~\ref{heating_soliton} shows the evolution of the cumulative stellar number distribution $N(<r)$ for the case with an initial scale radius $a=0.7$ kpc. Here $N(<r)$ denotes the number of stellar particles within a radius $r$ from the point of maximum stellar number density. The black dashed line represents the initial distribution, while the colored solid lines show the results after $10$ Gyr of evolution. In all three cases, the stellar distribution becomes more extended as a result of dynamical heating. The single-field and two-field cases exhibit very similar degrees of expansion, whereas the expansion in the four-field case is noticeably weaker than in the other two.

The upper-right panel of Fig.~\ref{heating_soliton} compares the radial profiles of the velocity dispersion along the $z$-direction before and after the evolution, with the dashed and solid lines denoting the initial and final states, respectively. Because the three halo models have different initial density profiles, their initial stellar velocity-dispersion profiles are correspondingly different. However, there is no clear evidence that the dynamical heating becomes weaker as the number of field components increases. Instead, the increase in the velocity dispersion is even slightly larger in the two-field and four-field models than in the single-field case.

Taken together, the upper two panels of Fig.~\ref{heating_soliton} indicate that the dynamical heating does not decrease with an increasing number of field components in the same way as the power spectrum or the two-point correlation function. This is because the later two parameters quantify only the strength of the granule-induced density fluctuations. In the present setup, however, the oscillations and random walks of the central core also provide a significant source of dynamical heating, thereby obscuring the expected scaling with the number of field components.

The lower two panels of Fig.~\ref{heating_soliton} show the corresponding results for an initial stellar scale radius of $a=2$ kpc. Compared with the $a=0.7$ kpc case, the stellar system is initially distributed over a much larger spatial extent, so the contribution of the central core to the dynamical heating is expected to be less important. Consequently, the accumulated stellar number distribution exhibits the anticipated trend more clearly, with the heating strength decreasing from the single-field model to the two-field model and then to the four-field model. However, the corresponding increase in the velocity dispersion still does not exhibit such a clear trend.

\begin{figure}[htbp]
    \centering    
    \includegraphics[width=\linewidth]{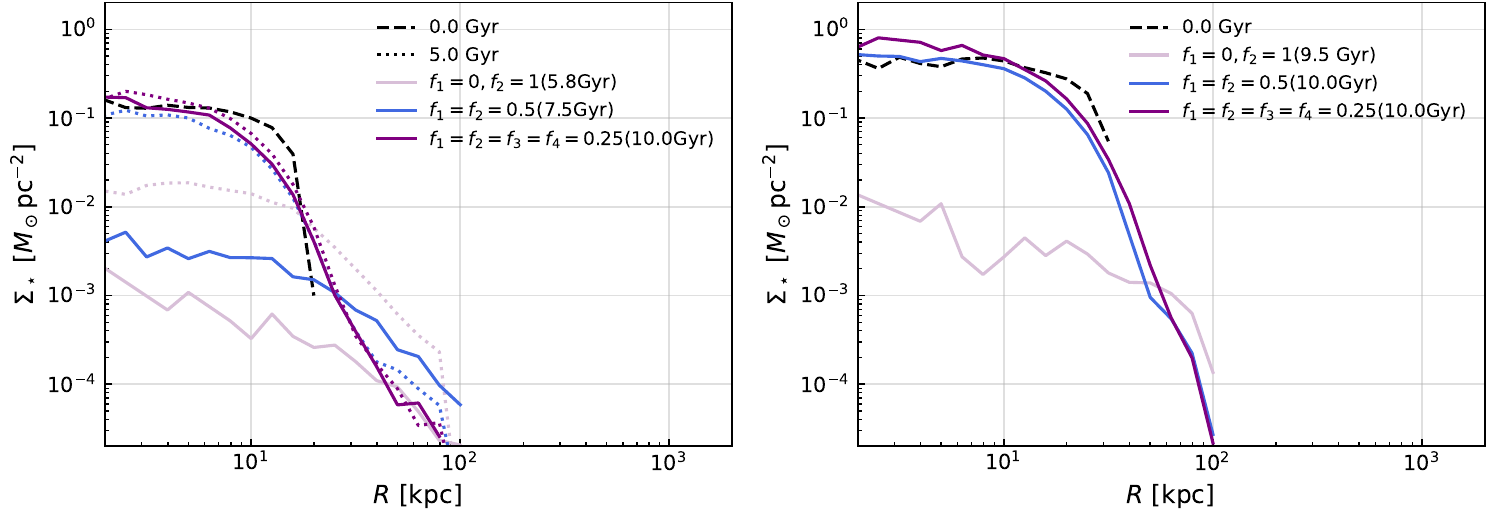}
    \caption{Evolution of the stellar surface density for stellar systems with a total mass of $100\, M_\odot$and an initial half-light radius of $15$ pc (left panel), and with a total mass of $1000\, M_\odot$and an initial half-light radius of $27$ pc (right panel).}
    \label{heating_granules}
\end{figure} 

To isolate the dynamical heating produced by granule fluctuations and eliminate the contribution from the motion of the central core, we consider a self-gravitating stellar system whose center is artificially fixed at a distance of $5$ kpc from the halo center. Since the system has a characteristic size of only $O(10)$ pc, the coherent oscillations and random walks of the central core have a negligible direct influence on its internal dynamics. We consider two sets of stellar-system parameters. The first has a total stellar mass of $100\, M_\odot$ and a half-light radius of $15$ pc, while the second has a total stellar mass of $ 1000 \, M_\odot$ and a half-light radius of $27$ pc. In both cases, the stellar system is represented by $10^4$ equal-mass particles. These parameters are chosen purely for illustrative purposes and are not intended to represent any particular observed stellar system.

Because the stellar systems considered here are extremely compact, simultaneously resolving their internal dynamics and the FDM wave dynamics would require prohibitively high spatial and temporal resolution. We therefore adopt the two-stage simulation strategy proposed in Ref.~\cite{yang2026collapseversusdisruptionfate}. In the first stage, the multifield FDM halo is evolved independently, during which the tidal tensor $\nabla\nabla V$ at a position $5$ kpc from the halo center is recorded as a function of time. In the second stage, the internal evolution of the stellar system is followed with the direct $N$-body code \textsc{PeTar} \cite{Wang_2020}. Throughout this evolution, the time-dependent tidal acceleration derived from the recorded tidal tensor is applied to the stellar particles. Further details of the simulation procedure can be found in Ref.~\cite{yang2026collapseversusdisruptionfate}.

Fig.~\ref{heating_granules} illustrates the evolution of the stellar surface density for the two sets of stellar-system parameters. The left panels correspond to the model with a total stellar mass of $100\, M_\odot$ and an initial half-light radius of $15$ pc. In the single-field model, the stellar system is almost completely disrupted by $\sim5.8$ Gyr. In contrast, disruption occurs only at $\sim 7.5$ Gyr in the two-field model, while the four-field model shows no clear sign of disruption throughout the entire $10$ Gyr simulation. The right panels show the corresponding results for the stellar system with a total mass of $1000\, M_\odot$ and an initial half-light radius of $27$ pc. In this case, only the stellar systems in the single-field model is disrupted, at approximately $9.5$ Gyr, whereas the stellar systems in the other two models remain bound throughout the simulation. These results provide clear evidence that the dynamical heating induced by granule fluctuations becomes progressively weaker as the number of field components increases.

\section{Conclusions\label{Sec_7}}
In this work, we have performed a systematic numerical study of the dynamical properties of wave structures in multifield FDM halos. By evolving halos formed through soliton mergers, we have investigated how the presence of multiple fields modifies the behavior of central cores, the statistical properties of granules, and the resulting impact on embedded stellar systems. Our main findings are summarized as follows:

\begin{enumerate}

\item The central cores in multifield FDM halos exhibit qualitatively different oscillation patterns from those in the single-field case. The density oscillation spectra develop multiple characteristic peaks, accompanied by a shift toward higher frequencies. This reflects the presence of additional dynamical modes introduced by the multiple field components and demonstrates that multifield systems possess a richer internal structure than their single-field counterparts.

\item The centers of individual field components experience stochastic motions driven by the evolving gravitational potential. Despite differences in field abundance, the cores of different components remain strongly correlated and undergo nearly synchronized random walks, consistent with previous studies \cite{Huang_2023}. We also find that components with smaller mass fractions generally exhibit larger random-walk amplitudes, suggesting that their motions are strongly perturbed by the gravitational potential of the dominant components.

\item The statistical properties of granule fluctuations are strongly affected by the number of fields. Through measurements of both the density power spectrum and the two-point correlation function, we show that the amplitude of granule density fluctuations decreases approximately as the inverse of the number of fields, consistent with previous studies \cite{Gosenca_2023}. In contrast, varying the fractional abundance among individual components produces only a weak modification to the overall fluctuation strength over the explored parameter range. This suggests that the suppression of granule fluctuations is primarily controlled by the multiplicity of fields rather than by their specific mass partition.

\item The reduced granule fluctuations in multifield FDM halos lead to weaker dynamical heating of embedded stellar systems. Using coupled FDM and $N$-body simulations, we demonstrate that the heating efficiency induced by granules decreases as the number of fields increases. However, when the contribution from the oscillating central core is included, the dependence on the number of fields nearly vanishes, suggesting that core-driven gravitational fluctuations can remain an important source of stellar heating even when granule fluctuations are suppressed.

\end{enumerate}

Overall, our results establish a comprehensive picture of the dynamical evolution of wave structures in multifield FDM halos. The additional degrees of freedom introduced by multiple fields substantially modify both the coherent core dynamics and the incoherent granule fluctuations, leading to potentially important consequences for the astrophysical signatures of FDM. 

\acknowledgments
This work is supported by the National Natural Science Foundation of China under Grants No. 12447105 and No. 12575113.

\appendix

\bibliography{Refs}

\end{document}